%

\magnification 1200
\baselineskip 16 pt

\topskip .7 in
\leftskip .25 in
\rightskip .25 in
\vsize 7.2 in
\vglue .25 in

%
\def\eq#1{ \eqno({#1}) \qquad }
\def\ea{{\it et al\/}}
\def\ie{{\it i.e.\ }}
\def\ni{\noindent}
\def\Par{\par \vskip .2cm}

\font\bigggfnt=cmr10 scaled \magstep 3
 2
\font\bigfnt=cmr10 scaled \magstep  1

\def\NY{N\'u\~nez-Y\'epez }
\def\fis{\left (\dot \phi \over \phi \right ) }

 \newbox\Ancha
 \def\gros#1{{\setbox\Ancha=\hbox{$#1$}
   \kern-.025em\copy\Ancha\kern-\wd\Ancha
   \kern.05em\copy\Ancha\kern-\wd\Ancha
   \kern-.025em\raise.0433em\box\Ancha}}


\noindent
{\bigggfnt Singularities and isotropy in a Brans-Dicke\par\vskip 4 pt
\ni Bianchi-type VII  universe} \Par 

\vskip 12 pt

\noindent
{\bigfnt H N N\'u\~nez-Y\'epez}{\footnote{\dag}{\rm On sabbatical leave from Departamento de F\'isica, UAM-Iztapalapa, \par e-mail: nyhn@xanum.uam.mx}} \Par

\noindent
{Instituto de F\'{\i}sica, Benem\'erita Universidad Aut\'onoma de Puebla, A\-par\-tado Pos\-tal J-48, C P 72570,  Puebla, M\'exico} \Par
\vskip 14 pt

\centerline {\bigfnt Abstract.} 
We study the dynamical effects in the scale factors due to the scalar $\phi$-field at the early stages of a supposedly anisotropic Universe expansion in connection with the problem of the initial singularity in the scalar-tensor cosmology of Jordan-Brans-Dicke. This is done by considering the behaviour of the general analytical solutions for the homogeneous model of Bianchi type VII in the vacuum case. We conclude that the Bianchi-VII$_0$ model shows an isotropic expansion and that its only physical solution is equivalent to a Friedman-Robertson-Walker spacetime whose evolution begins in a singularity and ends in another; moreover, we obtain that the general Bianchi-VII$_h$ (with $h\neq 0$) models show strong curvature singularities that produces a complete collapse of the homogeinity surfaces to a 2-plane, to a string-like  one-dimensional manifold, or to a single point. \Par
\vskip 12 pt


\noindent{ PACS number(s)}: 04.20.Jb \Par 

\vfill
\eject

\noindent
{\S (1) Introduction} 

In the Jordan-Brans-Dicke scalar-tensor theory of gravity (JBD)[1--2](Jordan 1959, Brans and Dicke 1961), a massless scalar $\phi$-field is introduced in addition to the pseudo-Riemannian metric of the spacetime $R_4$ occurring in Einstein general relativity (GR); supposedly, this long range field is generated by the whole of matter in the Universe according to Mach's principle [3](Dicke 1964); furthermore, the inclusion of this field, allows Dirac's idea about the secular variation of the gravitational constant $G$ to occur in JBD cosmology. The scalar field in JBD acts as an additional effective source of $R_4$ geometry, and it is coupled to the tensorial degrees of freedom of the theory by a constant parameter $\omega$ [4](Ruban and Finkelstein 1975). The value of $\omega$ can be estimated from astronomical observations as $|\omega| \simeq$ 500 to be in accord with current observations, but, as the GR limit of the theory practically starts at such value, the theory does not seem to predict anything too different from GR with enough observational evidence to support it. \par

However, despite what we have just said, there is a renewed interest in JBD (and other scalar tensor theories), mainly because of superstring theories which lead naturally to a dilaton theory of gravity where scalar fields are mandatory, and also due to the emergence of extended inflation models and pre-big bang cosmologies where scalar fields can provide solutions to some of the problems of inflation [5--6](Steinhard 1993, Gasperini and Veneziano 1994). This comes about since the JBD action functional already includes a string sector where the dilaton field $\phi_D$ can be suitably related with the JBD scalar field $\phi \propto \exp(-\phi_D)$. The important role of the $\phi$-field of JBD would especially occur at the strongly relativistic stages of the Universe expansion in view of the important role that the so-called scale factor duality plays in the pre-big bang scenario [6,7](Gasperini and Veneziano 1994, Clancy \ea\ 1998). Thence the importance of studying Bianchi universes in JBD cosmology. In this work we analyse the Bianchi type VII vacuum universes and show that these exhibit curvature singularities which make their surfaces of homogeinity collapse to a plane, to a string or, even, to a point, and that, despite the suppposed anisotropic behaviour of a Bianchi type VII universe, a Bianchi VII$_0$ evolves isotropically and becomes essentially equivalent to a Friedman-Roberson-Walker universe starting from a singularity and ending in another. \Par 

\noindent
{\S (2) Vacuum Bianchi-type VII equations}

Let us write for the line element of the spacetime, using signatu\-re $+2$  and natural units $c=G=1$,

$$ ds^2 \, = \, -dt^2 \, + \, h_{ij}(t) \, {\gros \omega}^i \, {\gros
\omega}^j,
\eq{1} $$

\noindent
where the $h_{ij}$ is the metric on the surface of homogeinity, $t$ is the synchronous time, ${\gros \omega}^i$ are the one-forms [8](Ryan y Shepley 1975) expressing the properties of the 3-space; the specific one-forms appropriate for the homogeneous but anisotropic Bianchi VII model are

$$ \eqalign{
{\gros \omega}^1 =& a_1 \left( (\eta  - k \nu) dy - \nu dz \right), \cr
{\gros \omega}^2 =& a_2 \left( \nu dy - (\eta  + k \nu) dz \right), \cr
{\gros \omega}^3 =& a_3 dx, \cr
{\gros \omega}^4 =& dt,
} 
\eq{2} $$

\noindent 
where $\eta \,=\, \exp (-kx) \cos(M x)$, $\nu \,=\, (-M^{-1}) \exp (-kx) \sin(M x)$, $k \,=\, h/2$ and $M \,=\, (1-k^2)^{1/2}\,$; the parameter $h$ distinguish the case of the special model VII$_0$ ($h=0$) from the generic  model VII$_h$ ($h \neq 0$). \par

If we now assume a vacuum model and insert the line element (1), with the forms (2), into the JBD field equations in a vacuum, we get

$$ {\eqalign {
{d^2 \over dt^2}(\ln a_i) &+
{d\over dt} (\ln a_i) {d\over dt} (\ln a_1 a_2 a_3) +
{d\over dt} (\ln a_i)                                  \fis \cr 
&+ {\cal A}_i a_1^{-2} + {\cal E}_i \beta_2 + {\cal F}_i \beta_3 = 0,  }} \quad i={ 1,2,3}
 \eq{3} $$

\noindent 
where one of the coordinates has been chosen as the synchronous time $t$. We additionally have what we have called the constriction equation, coming from off-diagonal terms in the JBD field equations,

$$ \eqalign {
&{d\over dt}(\ln a_1) {d\over dt}(\ln a_2) +
 {d\over dt}(\ln a_1) {d\over dt}(\ln a_3) +
 {d\over dt}(\ln a_2) {d\over dt}(\ln a_3)  \cr 
& + {d\over dt}(\ln a_1 a_2 a_3) \fis - {\omega \over 2} \fis^2 +
    {\cal A}_4 a_1^{-2} + {\cal E}_4 \beta_2 + {\cal F}_4 \beta_3 = 0,
}
 \eq{4} $$

\noindent 
this is an equation of Raychaudhuri-type [9]; finally, the scalar field
comply with

$$ {d\over dt} \left([a_1 a_2 a_3] {d\phi\over dt} \right) = 0,
 \eq{5} $$

\noindent 
where $\beta_i \equiv (a_i  / (2 a_j a_k))^2$ and the indexes $i,j,k$ are to be taken in cyclic order of 1,2,3. Equations (3), (4) and (5) are written in the standard form we introduced previously for solving the Bianchi-models [9](Chauvet \ea\ 1992); the specific values for the constants appearing in them are: ${\cal A}_1 = 4 M^2 - (5/2)$, ${\cal
A}_2 = {\cal A}_1 - 2 $, ${\cal A}_3 = 0$, ${\cal A}_4 = {\cal A}_1 - 1$, ${\cal E}_1 = {\cal E}_3 = {\cal F}_1 = {\cal F}_2 = -2$, ${\cal E}_2 = {\cal F}_3 = 2$, ${\cal E}_4 = {\cal F}_4 = -1$, and $M^2 = 1 - (h^2 /4)$; compare with Table 1 in [9](Chauvet \ea\ 1992). Notice that according to these relationships, $h$ must be restricted to be $|h|\leq 2$ (this can also be seen from the differential forms in (2)). Also, the equations for the Bianchi VII$_0$ model can be particularized from the general ones, just by taking $h=0$.

Most Bianchi models, as consequence of the off-diagonal contributions to the field equations, lead to additional relationships between their scale factors; in this case, we have the following two additional relationships

$$ {d\over dt}(\ln a_1) - {d\over dt}(\ln a_2) = 0,   
\eq{6} $$

\vskip -8 pt

$$ {h a_2 \over 2  a_1^2  a_3}  =  0. 
 \eq{7} $$

\noindent
Equation (6) always provides an additional relationship between the two scale factors, $a_1$ and $a_2$, not mattering what the value of the 
$h$-parameter; but, on the other hand, equation (7) becomes just a trivial identity when $h=0$, not restricting in any way the values of the scale factors. The case $h \neq 0$ and some of its consequences are analysed in subsection 3.2. \par

\centerline {2.1 {\sl Scaling the scalar field}}

From equation (5), we easily obtain

$$ a_1 \, a_2 \, a_3 \, {d\phi\over dt} \,=\, \phi_0,
 \eq{8} $$

\noindent
where $\phi_0$ is an integration constant; thus we can introduce the scaled $\phi$-field, $\Phi$, as $\Phi \,\equiv\, \phi / \phi_0$, which  coincides with the so-called intrinsic time [10](Carretero-Gonz\'alez \ea\ 1994). From equation (8), we easily get

$$ \partial_t \,=\, (a_1 \, a_2 \, a_3)^{-1} \, \partial_{\Phi}.
 \eq{9}  $$

\noindent
This shows that $\Phi$ is a monotonic function of the synchronous time $t$; $\Phi$ can hence be used also as a time reparametrization useful for solving equations (3). In fact, $\Phi$ has been found useful for analysing the Bianchi vacuum models in several situations [9--12](Chauvet \ea 1991, 1992, Carretero-Gonz\'alez \ea\ 1994, N\'u\~nez-Y\'epez 1995). For the sake of convenience, let us introduce the notation $(\,)^{\prime} \,\equiv\, \partial_{\Phi}$ and, defining the Hubble expansion rates as $H_i \,\equiv\, (\ln a_i)^{\prime}$, the reparametrized field equations become

$$
H_i^{\prime} +\, {H_i \over \Phi} +\, {\cal J}_i\,a_2^4 \,+\, {\cal K}_i\,a_3^4 \,+\, {\cal N}_i\,a_2^2\, a_3^2 \,=\, 0,
\quad i={ 1,2,3}
 \eq{10} $$

\noindent
and the constriction equation becomes

$$ \eqalign {
H_1\,H_2 \,&+\, H_1\,H_3 \,+\, H_2\,H_3 \,+\, {(\ln a_1 \, a_2 \,
a_3)^{\prime} \over \Phi} \,-\, {\omega \over 2 \, \Phi^2} \cr
&+\, {\cal J}_4\,a_2^4 \,+\, {\cal K}_4\,a_3^4 \,+\, {\cal N}_4\,
a_2^2\,a_3^2
\,=\, 0,
}
\eq{11} $$

\noindent
the specific values for the constants appearing in equations (10) and (11) are combinations of the constants previously used: ${\cal J}_1 = {\cal J}_3 = {\cal K}_1 = {\cal K}_2 = -1/2$, ${\cal J}_2 = {\cal K}_3 = 1/2$, ${\cal J}_4 = {\cal K}_4 = -1/4$, ${\cal N}_1 = 4 M^2 - (5/2)$, ${\cal N}_2 = {\cal N}_1 - 2$, ${\cal N}_3 = 0$, ${\cal N}_4 = {\cal N}_1 - 1$. The specific form chosen to write the equations and the parameters just emphasizes the relationship with our previous work [9,11,12] (Chauvet \ea\ 1991, 1992, \NY\ 1995) on exact solutions for vacuum Bianchi models in JBD. \Par

\noindent
{\S(3) Solutions for the vacuum Bianchi type VII universes}

For solving the equations of the Bianchi-type VII model, we found convenient to address separately the specific VII$_0$ and the generic VII$_h$ Bianchi models. The exact solutions of the next subsections are obtained using essentially the method in Chauvet \ea\ [9](1992, Appendix).

\centerline {3.1 {\sl Bianchi type VII$_0$}}

In the reparametrized formulation of the equations for the anisotropic homogeneous metric of Bianchi type VII$_0$, solutions can be obtained for the case of a Bianchi-type VII$_0$; the specific solution depends on the sign of the quantity $\Delta \equiv - 4 ({\cal B} + 1/4)$ where 

$${\cal B}=\Phi^2 H_1^2-{\Phi\over 2} H_1- {\Phi^2 \over 2}H_1', \eq{12} $$ 

\noindent 
is a constant, \ie\ is a first integral of the reescaled system (3) and (4) that depends on the Hubble expansion rates; in this way, we can find that out of the possible solutions of equations (10), the only physically plausible is the one corresponding to the case $\Delta < 0$ (the $\Delta > 0$ or $\Delta = 0$ cases can be seen to led to negative or even complex scale factors); for some details see [12](Chauvet \ea\ 1992). The only physical solution can be explicitly written as

$$ a_1(\Phi) = \left( 4\, {\cal B} + 1 \over {c_0}^4 \right)^{1/4} \,
\left\{ \Phi \cosh \left[ - \sqrt{ ({\cal B} + 1/4)} \ln (f \Phi^2) \right] \right\} ^{-1/2},
\eq{13} $$

\noindent
where $c_0$ and $f$ are positive integration constants. The other two scale factors can be easily obtained from $a_1(\Phi)$, as follows from (6) and (10), they are 

$$ a_2(\Phi) = c_0 \, a_1(\Phi) , 
\eq{14}$$

\vskip - 8 pt

$$ a_3(\Phi) =\ 2^{-1/2} c_0 a_1(\Phi); 
\eq{15} $$

\noindent
the three scale factors are proportional to each other. This means that the Bianchi-VII$_0$ model, despite what we could have anticipated, shows an {\sl isotropic expansion}; it also implies that the shear, vorticity and acceleration of the reference congruence all vanish. The vacuum Bianchi-VII$_0$ JBD universe behaves as a Friedmann-Roberson-Walker (FRW) space-time---in a way, this is not totally surprising since Bianchi cosmologies correspond to the simplest deviations from a FRW environment. The local volumen on the surface of homogeneity is then

$$ V \,=\, {c_0^2 \, (a_1)^3 \over \sqrt{2}}.
\eq{16} $$

The constriction equation implies the following relationship in our case 

$$ 12 \, \left( {\cal B} \, + \, {1 \over 4} \right) - (3 \, + \, 2 \, \omega) = 0;
\eq{17} $$

\noindent
we notice that to have meaningful solutions the coupling parameter has to be restricted to $\omega > -3/2$. Figure 1 shows the evolution of the scale factors of the model as a function of  the intrinsic time (or reescaled $\phi$-field) $\Phi$. \par

Using equations (9) and (16) we can obtain the dependence of $\Phi$ on
$t$, as follows

$$ t = { (4 {\cal B} + 1)^{3/4} \over \sqrt{2} \, {c_0}} \int \{\Phi \cosh [ \sqrt{{\cal B} + {1/ 4}} \, \ln(f \Phi^2)]\}^{-3/2} d \Phi,
\eq{18} $$

\noindent
although we can obtain $t$ as a function of $\Phi$, as in (18), we cannot invert it to obtain explicitly $\Phi$ as a function of $t$.  Nevertheless, figure 2, exhibits the dependence of the field $\Phi$ on $t$, showing the enormous change that occurs in $\Phi$ over a very small span of $t$ values. This also shows that, asymptotically, $\Phi$ grows without bound and that as $t\to 0$, $\Phi$ vanishes too; notice that these conclusions can be applied with no changes to the scalar field $\phi$, excepting when $\phi_0=0$. We can now relate the behaviour of $\Phi$ with that of $a_1$; since, as figure 1 shows, the universe begins with a singularity and then, as $\phi$ grows, the universe expands as $\sim t^{\alpha}$, where $\alpha \equiv 3\sqrt{{\cal B} +1/4}-3/2$ ---it thus does not really expand unless $ {\cal B} > 0$--- rapidly reaching a maximum volume $V_{\hbox{max}}$ that can be easily calculated from (13); and then, as $t \to \infty$, it shrinks rapidly as $\sim t^{-(\alpha /3 + 1/2)}$ until it reaches again a singularity; see figures 1 and 3. \par

\centerline {3.2 {\sl Bianchi type VII$_h$}} 

From equations (6) and (7) several relations can be obtained for the
scale factors irrespective of the value of $h \neq 0$; from (6) we get (14) (the same equation than in the case $h=0$). From equation (7), we might get $a_2 = 0$, implying that $a_1 = 0$ and $a_3$ can take any value; then the model collapses into essentially a spatial 
one-dimensional manifold: \ie\ a string. Other possible options allowed by equation (7), imply that $a_2 \, a_3 = \infty$, thus the spatial 
3-surfaces of homogeinity are seen to collapse into 2-surfaces. The important point here is that for any choice of values for the scale factors in equation (7), the model is found always to spatially collapse. Any Bianchi-type VII$_h$ ($h \neq 0$) universe is thus highly singular. \par

As we have been able to get the scale factors for the especific cases addressed in this paper, we have obtained exact vacuum solutions for the Bianchi VII models. The important conclusion is that all solutions found  for the generic Bianchi VII model with $h=0$, and $h \neq 0$---with the restriction $|h| \leq 2$--- describe collapsing singular universes. The behaviour of the scale factors is exhibitted in figure 1. \Par

\noindent
{\S(4) Curvature singularities in the Bianchi-type VII universes} 

In this section we study the curvature singularities present on the Bianchi-VII universes, though some very specific singularities were discussed in section 3.2. \par

We say that a universe is singular if the value of the Ricci scalar $R = g^{ab} R_{ab}$ along a congruence of the geodesics $|R| \to \infty$ whereas the associated affine parameter tends to a finite value. If this happens, then we say that there exists a curvature singularity and thus that the universe is also singular. Notice also that, due to the choice we made for the $\omega$ parameter, here the singularity occurs when $R \to - \infty$ (figure 3) rather than the other way round. The problem is basically how to choose an appropriate geodesic congruence. For our Bianchi VII models, we have chosen as the proper congruences the world lines of test observers (time-like geodesics) whose affine parameter is the synchronous time $t$, this is the simplest choice we could found. Thence, the scalar curvature is  found to be

$$ R \,=\, 2 \left( {\ddot a_1 \over a_1} + {\ddot a_2 \over a_2} +
{\ddot a_3 \over a_3} + {\ddot \phi \over \phi} + {\omega \over 2}
\left( {\dot \phi \over \phi}\right)^2 \right).
\eq{19} $$

\noindent
Using only equations (5), (10) and (11) we can rewrite $R$ for the vacuum Bianchi-VII model in terms of the coupling parameter $\omega$ and the scalar field $\phi$, as follows

$$ R \,=\, - \, \omega \, \left( {\dot \phi \over \phi} \right)^2
     \,=\, - \, \omega \, \left( {1 \over a_1 \, a_2 \, a_3 \, \Phi}
\right)^2;
\eq{20} $$

\noindent
it is important to notice that expressions (19) and (20) do not depend on the $h$-value, they are valid for all the Bianchi-type VII models we are discussing. We have expressed the scalar curvature in two different ways, equations (19) or (20), both are important because they exhibit  the explicit dependence of $R$ on the scale factors and $\phi$ and its time derivatives, or on the coupling paramenter $\omega$ and the scale factors (20) and, besides, they have both a certain pleasant simplicity. For $R \to \infty$ in (20), all that is needed is that at least one of the scale factors ($a_1$, $a_2$, or $a_3$) or the reescaled scalar field $\Phi$, or just the scalar field $\phi$, vanish at a finite value of the synchronous time $t$. Notice also that we can regard the evolution of the Bianchi-type VII universe ---equations (3) and (4)--- as driven by  the curvature $R$, this is especially true near the singularity. \par
 
Notice that the plot of $t$ against $\Phi$ in figure 2, which does not depend on $h$ in any way, shows the rather small range of $\Phi$-values in which the expansion of the model universe VII$_0$ occurs, as we can see comparing with figure 1; this also corresponds to the region free of singularities in the model (VII$_0$), as can be seen on comparing with figure 3. On comparing figures 1 and 2, the role of the curvature scalar in governing the expansion can be qualitatively described as follows: a strong curvature prevents expansion; it is only when curvature is small that the full extent of expansion is reached but, as soon as the curvature is large in magnitude again, contraction sets in. This qualitative behaviour is in accord with current ideas [7, 13].

In  the general VII$_h$ JBD model, the universe is always singular, as can be easily seen from equation (7) and the discussion in section 3.2; in this sense, we say that the model VII$_h$ is completely singular. \Par

\noindent
{\S(5) Concluding remarks} 

The supposed homogeneous and anisotropic Bianchi VII model in fact shows an isotropic expansion  in the case in which $h=0$. On the other hand, this article shows  that the dynamics of the early stages of the expansion in the specific VII$_0$ JBD model depends on just one of the scale
factors (that, here, we choose as $a_1$). We have also obtained the dependence of the three scale factors $a_i$ on the reescaled field $\Phi$. As we have concluded that the three scale factors are proportional to each other, the expansion is necessarily isotropic and not anisotropic, as it is  usually  assumed to be in this model. In fact, we have shown that a Bianchi-VII$_0$ JBD vacuum universe is basically equivalent to a FRW-spacetime. But the important point is that this  shows that, even with supposedly anisotropic models, the inclusion of a scalar field can sometimes isotropize the behaviour, which thus offers the possibility of coordinating it with the observed isotropic properties of our Universe. Moreover, this feature does not seem to depend on $\omega$ in any way.

For the case of the VII$_h$ vacuum model, we can  say that, not mattering the choice   made in equation (7), the universe always collapses. According to the discussion in section 3.2, the universe may spatially collapse into a plane or into a one dimensional object or, even, into a  singular point; this model universe always collapses to a permanent singularity. Such behaviour may make this  model of little interest from a factual point of view unless one is interested in singular behaviour. In this respect, the dependence of the universe dynamics, especially near the singularity, on the curvature scalar is worth pinpointing.\Par

\centerline { Acknowledgements.} 

This work has been partially supported by CONACyT (grant 1343P-E9607). 
\vfill
\eject

\centerline { References}\par

\item [1 ] Jordan, P.\ (1959).  {\sl Z.\ Phys.}, {\bf 157}, {112}.

\item [2 ] Brans, C.\ and Dicke, R.\ H.\ (1961).  {\sl Phys.\ Rev.}, {\bf 124}, {925}.

\item [3 ] Dicke, R.\ H.\ (1964).\ {\it Gravitation and Relativity}, (eds.\ {Chiu, H.\ and Hoffman, W.),  (W.\ A.\ Benjamin, Inc., New York and Amsterdam). }

\item [4 ] {Ruban, V.\ A.\ and Finkelstein, A.\ M.\ (1975).} {\sl Gen.\ Rel.\ Grav.}, {\bf 6} {601}.

\item [5 ] {Steinhard, P.\ J.\ (1993)}. {\sl Class.\ Quantum Grav.}, {\bf 10}, {S33}.

\item [6 ] {Gasperini, M.\ and Veneziano, G.\ (1994)}. {\sl Phys.\ Rev.\ D}, {\bf 50}, {2519}.

\item[7 ] {Clancy, D.\ Lidsey, J.\ E.\ and Tavakol, R.\ (1998)}. {\sl Class.\
Quantum Grav.\ }, {\bf 15}, {257}.

\item [8 ] {Ryan, M.\ P.\ and Shepley, L.\ C.\ (1975)}. {\it Homogeous relativistic cosmologies}, {(Princeton University Press, Princeton N.\ J.\ ) pp.\ 113--267}.

\item [9 ] {Chauvet, P.\ Cervantes-Cota, J.\ and N\'u\~nez-Y\'epez, H.\ N.\ (1992)}. {\sl Class.\
Quantum Grav.\ }, {\bf 9}, 1923.

\item [10 ] {Carrretero-Gonz\'alez, R.\  N\'u\~nez-Y\'epez, H.\ N.\ and Salas-Brito, A.\ L.\ (1994)}. {\sl Phys.\ Lett.\ A},  {\bf 188}, {48}.

\item [11 ] {Chauvet, P.\ N\'u\~nez-Y\'epez, H.\ N.\ and Salas-Brito, A.\ L.\ (1991)}. {\sl Astrophys.\ Space Sci.\ }, {\bf 178}, {165}.

\item [12 ] {N\'u\~nez-Y\'epez, H.\ N.\ (1995)}. {\it Soluciones exactas y caos en la cosmolog\'ia de Jordan, Brans y Dicke}, {Tesis doctoral, (Universidad Aut\'onoma Metropolitana, Mexico City)}

\item [13 ] {Turner, M.\ J.\ and Weinberg, E.\ J.\ (1997)}. {\sl Phys.\ Rev.\ D}, {\bf 56}, {4604}.

\vfill
\eject

\centerline{Figure Captions} \Par
\ni Figure{1}\par
The behaviour of the three scale factors $a_1$, $a_2$ and $a_3$ is shown in the Bianchi-type VII$_0$ universe. Notice that all scale factors are proportional to each other. This figure basically show graphs of equations (13), (14) and (15). The values used for the parameters are ${\cal B}=83.33$, $c_0=2$, $f=1$. \par\vskip 12 pt

\ni Figure {2}\par
The behaviour of the reescaled scalar field $\Phi$ is shown as a function of $t$. This graph can be also interpreted as the behaviour of the intrinsic time {\it versus} the synchronous time $t$, as follows from equation (18). Notice how the scalar field goes from a  certain finite value $\phi_0$ at $t\sim 0$ to an infinite asymptotic value in a very small $t$-lapse . \par\vskip 12 pt

\ni Figure {3}\par
The scalar curvature $R$ in the Bianchi-type VII$_0$ universe is shown in the range $-\infty<R<\infty$ {\it versus} $\Phi$ in the range $0<\Phi<\infty$. What we really graph here is $\arctan(R)$ against $\arctan(\Phi)$. Small $\Phi$-values  correspond to small $t$-values but  large values of $\Phi$ correspond to $t\,{}_\sim^>\, 3.5$, as can be seen in figure 2.

 \vfill
 \eject
  
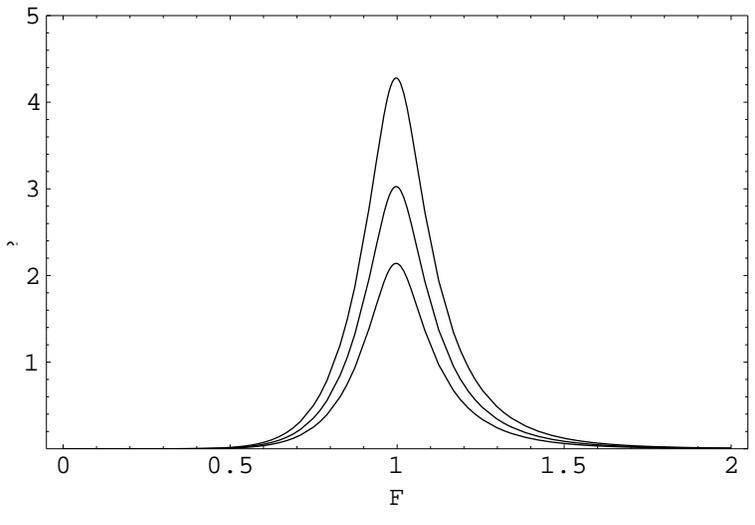
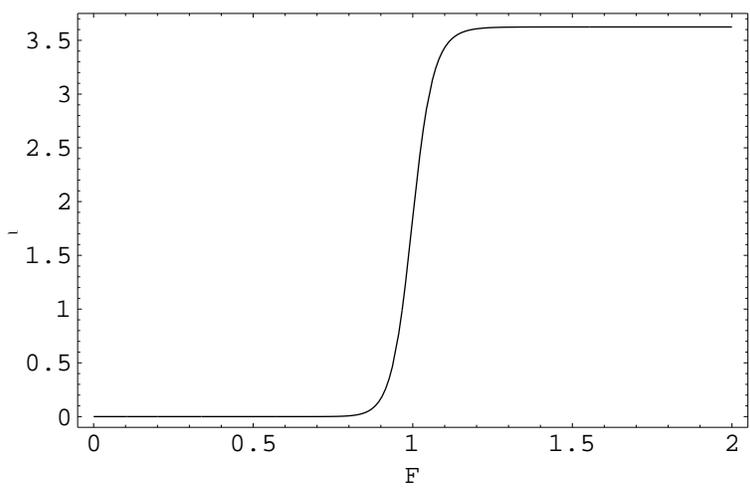
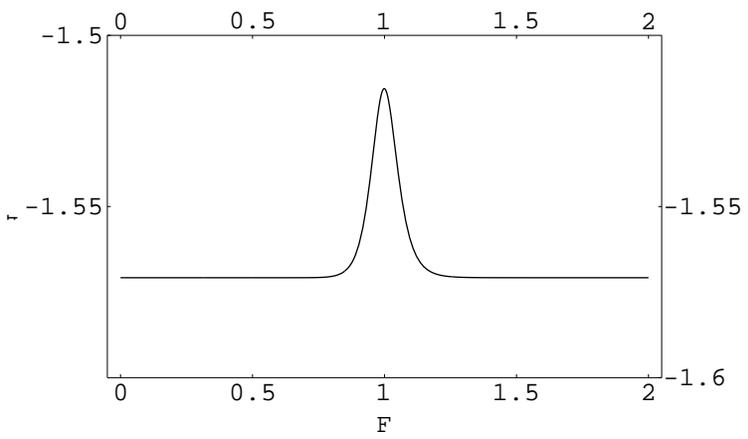
\end